\documentclass[a4paper,12pt]{article}

\usepackage[dvips]{graphicx}
\usepackage{amsmath}
\usepackage{amssymb}
\usepackage{subfigure}

\newcommand{\captive}[1]{\rule{5mm}{0mm}%
\begin{minipage}{0.9\textwidth}\caption[small]{#1}\end{minipage}}

\newcommand{\Pint}{-\hspace{-2.5ex}\int}

\def\dd{\mathrm{d}}
\def\Pd{\Pi^{\prime \prime}(s)}
\def\T{\mathcal{T}}
\def\Im{\mathrm{Im}}
\def\GeV{\mathrm{GeV}^2}

\def\qq{{}\langle \overline{q} q\rangle{}}

\def\mqq{{}\langle m_q \overline{q} q\rangle{}}
\def\agg{{}\langle a_s G^2\rangle{}}
\def\mss{{}\langle m_s \overline{s} s\rangle{}}
\def\mix{{}\langle g \overline{q} \sigma_{\mu \nu} G^{\mu \nu a}
\lambda^a q\rangle{}} 
\def\sig{{}\langle \overline{q} \sigma_{\mu \nu} \lambda^a q \overline{q}
\sigma^{\mu \nu} \lambda^a q\rangle{}}
\def\gam{{}\langle \overline{q} \gamma_{\mu} \lambda^a q \overline{q}
\gamma^{\mu} \lambda^a q\rangle{}}

\def\L{\log \frac{Q^2}{\mu^2}}
\def\LL{\log^2 \frac{Q^2}{\mu^2}}
\def\LLL{\log^3 \frac{Q^2}{\mu^2}}

\begin{document}
{\large
\hfill\vbox{\hbox{DCPT/01/108}
            \hbox{IPPP/01/54}}
\nopagebreak
}
\vspace{8mm}

\begin{center}
{\huge QCD
 Finite Energy Sum Rules}\\[3mm]
{\huge and the Isoscalar Scalar Mesons}\\[14mm]
{\Large S.~N.~Cherry$^{a}$ and M.~R.~Pennington$^{b}$}\\[8mm]
$^{a}$ \textit{Groupe de Physique Th\'eorique, IPN\\
Universit\'e de Paris-Sud,\\
F-91406 Orsay C\'edex, France.}\\[5mm]
$^{b}$ \textit{Institute for Particle Physics Phenomenology,\\
University of Durham,\\
Durham, DH1 3LE, U.K.}
\end{center}

\vspace{12mm}
\begin{abstract}
We apply QCD Finite Energy Sum Rules to the scalar-isoscalar current 
to determine the lightest $u \overline{u} + d \overline{d}$
meson in this channel.
We use `pinch-weights' to improve the reliability of the QCD
predictions and reduce the sensitivity to the cut-off $s_0$.
A decaying exponential is included in the weight function to allow us
to focus on the contribution from low mass states to the phenomenological
integral.
On the theoretical side we include OPE contributions up to dimension
six and a contribution due to instantons taken from the Instanton
Liquid Model.
Phenomenologically, we incorporate experimental data by using
a coupling scheme for the scalar current which links the vacuum polarisation to the $\pi
\pi$ scattering amplitude via the scalar form factor. 
We find that the sum rules are well saturated for certain instanton
parameters.
We conclude that the $f_0(400-1200)$ definitely contains a large
$u \overline{u} + d \overline{d}$ component, whereas the $f_0(980)$ most likely does not.
We are able to estimate the average light quark mass and find
$m_q(1~\GeV) = 5.2 \pm 0.6$ MeV.
\end{abstract}

\newpage

\baselineskip=6.0mm
\parskip=1.5mm

\section{Introduction}
\label{sec:intro}
Currently, experiment suggests that there are more scalar 
mesons than can be accommodated in a single quark-model nonet.
Many QCD motivated models predict the existence of non-$q
\overline{q}$ mesons such as glueballs~\cite{Fritzsch:1972},
hybrids~\cite{Horn:1978}, $q q \overline{q q}$~\cite{Jaffe:1977} and
$K \overline{K}$~\cite{Weinstein:1982} states.
It is tempting to say that these `spare' experimental states are the
unconventional mesons allowed by QCD but, with scalar glueball and
multiquark states expected to have masses comparable to the
conventional $q \overline{q}$ states, which \emph{are} the extra states?
In the work that follows we hope to shed some light on this question
by attempting to determine which of the 5 scalar-isoscalar mesons
currently listed in the Particle Data Group tables~\cite{Groom:2000} is the lightest $u
\overline{u} + d \overline{d}$ meson in this channel.
To address this question we will use the QCD sum rule technique.
The answer has importance for chiral symmetry breaking in QCD, see for instance~\cite{Scadron:1982,Meissner:1999}.

QCD sum rules are integral expressions that relate the hadronic and
partonic regimes, \emph{i.e.} the low energy world of resonances with
the high energy world of (tractable) QCD.
Since their conception over twenty years ago~\cite{Shifman:1979}, QCD sum
rules have become an established technique both for calculating
hadronic properties in channels, where the QCD expressions are under
control, and, conversely, estimating QCD parameters (such as the masses
of the quarks) in channels where there is good experimental
information. 

Using QCD sum rules in the scalar channel is not a new idea.
They were first applied to the scalar mesons
in~\cite{Reinders:1982} where Laplace sum rules were used (for a
recent review of Laplace Sum Rules see~\cite{Colangelo:2000}).
Here the phenomenological side was represented using the, now
standard, resonance + continuum ansatz, with resonances being
represented as $\delta$-functions and the continuum being calculated
entirely within perturbative QCD.
As only the Operator Product Expansion (OPE) terms, discussed in Sect.~\ref{sec:ope}, were
taken into account on the theory side, \emph{exact} mass degeneracy
between the lightest isoscalar and isovector was found, with
\mbox{$m_{f_0} = m_{a_0} = 1.00 \pm 0.03$~GeV}.
The $s \overline{s}$ state was predicted to have a mass of around
1.35~GeV.
These findings were supported by Bramon and Narison~\cite{Bramon:1989}, who used QCD
sum rules to calculate the couplings of the $f_0(980)$ and $a_0(980)$
to two photons.
Again modelling the resonances with $\delta$-functions the authors
concluded that a $q \overline{q}$ interpretation of these two mesons
could not be ruled out with the data then available.

The effects of going beyond the OPE were studied
in~\cite{Shuryak:1989}, where instantons (see
Sect.~\ref{sec:instantons}) were included on the theoretical side, but
not in the QCD continuum on the phenomenological side.
Once again the resonances were treated as $\delta$-functions and only
the isospin-1 channel was considered.
It was concluded that the lightest $q \overline{q}$ state in this
channel had a mass $\lesssim$~1~GeV.

For the light scalar mesons the zero-width approximation is not a
good one and the first attempt to go beyond it in a sum rule
investigation was by Elias \emph{et al}~\cite{Elias:1998}.
Here resonances were represented by Breit-Wigner formulae, which,
whilst better than a $\delta$-function, still do not adequately
describe the complex structure in the scalar sector.
A~further abstraction was introduced by replacing these Breit-Wigner
shapes with a Riemann sum of rectangular pulses.
Laplace sum rules were used, but in an integral form which requires the
calculation of perturbative expressions at rather low energy scales.
Again instantons were included, but not in the continuum.
In the isovector channel, the $a_0(980)$ was found to decouple from
the sum rule and the mass of the lightest quarkonium isovector was
consistent with the $a_0(1450)$.
In the isoscalar channel the conclusions were not so clear cut, it was
predicted that the lightest quarkonium here should have a width less
than half of its mass, and although the $f_0(980)$ could not be
ruled out as the main contributor a lighter state was preferred.
Similar conclusions were reached in~\cite{Shi:2000}.
This study used both Laplace and Finite Energy sum rules, again
requiring perturbative expressions to be evaluated at low energies.
The authors noted, that for a consistent treatment of the Borel sum
rules, the instanton effects should also be included in the QCD
continuum contribution.
The sum rules were dominated by an isoscalar with mass around 1~GeV and
an isovector with mass around 1.5~GeV, thus suggesting the $f_0(980)$
and $a_0(1450)$ as the lightest $q \overline{q}$ states in their
respective channels.
However, in the same work, a comparison to a more realistic resonance
shape predicted resonance parameters for the lightest quarkonium of $m
\sim 860$~MeV and $\Gamma \sim 340$~MeV.
These parameters are not consistent with the $f_0(980)$, but could
describe a Breit-Wigner fit to the $f_0(400-1200)$.

In~\cite{Maltman:1999}, Maltman used pinched weight finite energy sum rules 
to calculate the decay constants of the $a_0(980)$ and
$a_0(1450)$, which were then compared with the decay constant of the,
presumably $q \overline{q}$, $K_0^*(1430)$.
Instantons were included and all QCD integrals were carried out in the
complex plane, which improves the convergence of perturbative
expressions.
The isovector decay constants were found to be of comparable size,
suggesting a similar structure for both.
The author concluded that this favoured a Unitarised Quark
Model~\cite{VanBeveren:1986} scenario, where two physical hadrons can
arise from one `seed' state.

In the work that follows we will apply QCD Finite Energy Sum Rules introduced in Sect.~2 to
the scalar-isoscalar channel.
On the theoretical side we will include in Sects.~3 and 4 instanton effects as well as
the OPE terms up to dimension six and use the 
contour improvement prescription to increase the convergence of
perturbative expressions.
On the the phenomenological side in Sect.~5 we will incorporate
experimental data directly by relating the hadronic vacuum polarisation to the
$\pi \pi$ scattering amplitude via the pion's scalar form-factor.
The sum rules are evaluated in Sect.~6 and we present our conclusions in Sect.~7.

\baselineskip=6.0mm
\section{The Sum Rules}
\label{sec:rules}
The connection  between quarks and hadrons is through the vacuum
polarisation, $\Pi(s)$.
This is defined by the correlation function of two currents,
\emph{i.e.}
\begin{equation}
\Pi(s) = i \int \dd^4x \ e^{iqx}\ \langle 0|T\{J(x) J(0)\}|0 \rangle \ ,
\label{eq:corr}
\end{equation}
where $s = q^2 = -Q^2$ and the currents are chosen so as to select the
desired channel.
In this work we are considering the scalar-isoscalar mesons and so we
choose the Renormalisation Group invariant current
\begin{equation}
J(x) = \frac{m_q}{2} \left\{\overline{u}(x) u(x) + \overline{d}(x)
d(x) \right\} = \frac{\, m_q}{\sqrt{2}} \, \overline{n}(x) n(x)\ ,
\label{eq:current}
\end{equation}
where $m_q \equiv \tfrac{1}{2}(m_u + m_d)$ and we use the symbol $n$
to denote an effective light-quark.
The second equality follows because we ignore the two-gluon
intermediate state, whose contribution is $\mathcal{O}(m_q^2)$ and so
can be safely neglected.
\begin{figure}[ht]
\begin{center}
\includegraphics[height=0.3\textheight]{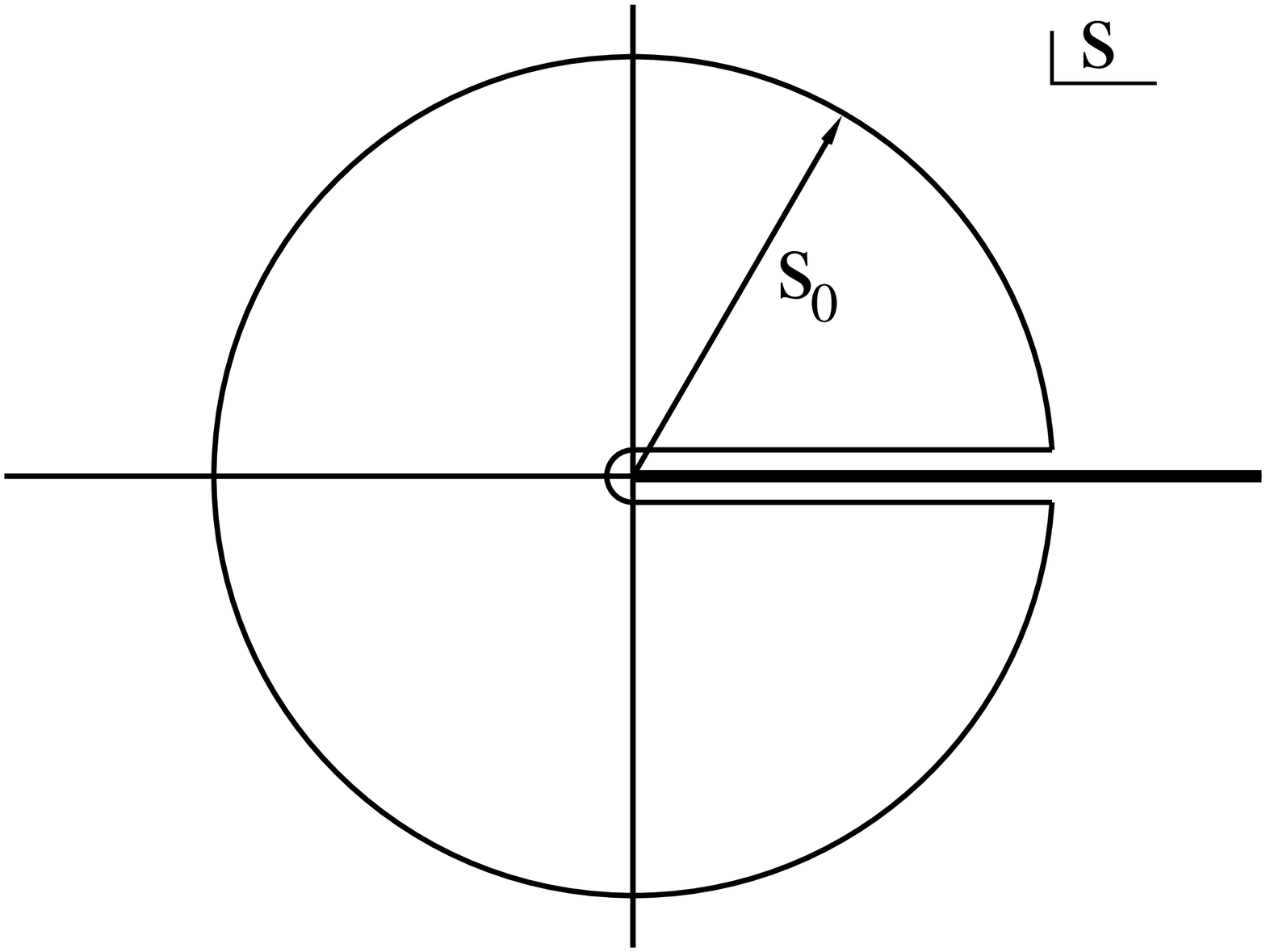}
\captive{The complex
$s$-plane showing the so called `Pacman' contour which is used to
derive the FESRs.
\label{fig:pacman}}
\end{center}
\end{figure}

$\Pi(s)$ is analytic everywhere in the complex $s-$plane except along
the time-like axis, so for any weight function, $w(s)$, which is real
for all real $s$ and analytic within the `Pacman' contour shown in
Fig.~\ref{fig:pacman} we can write 
\begin{equation}
\frac{1}{\pi} \int_0^{s_0} w(s) \, \mathrm{Im}\Pi(s) \, \dd s =
-\frac{1}{2 \pi i} \oint_{|s|=s_0} \!\!\!\! w(s) \, \Pi(s) \, \dd s \ . 
\label{eq:fesr}
\end{equation}

On the left-hand side of~Eq.~(\ref{eq:fesr}) we have an integral in the
physical region, and  here we choose to describe the integrand using
hadronic physics (either with experimental data or a phenomenological
parameterisation).
On the right-hand side we have an integral in the non-physical region
(complex $s$), and here we use field theory (\emph{i.e}. QCD) to
calculate the integrand.
The assumption that~Eq.~(\ref{eq:fesr}) holds for a range of values of
$s_0$ is a statement of global quark-hadron duality, \emph{i.e.} the
integral of an hadronic expression is equal to the integral of an
expression written in terms of partonic degrees of freedom.

Early sum rule work used positive integer powers of $s$ as weight
functions.
In this work we wish to investigate the lightest meson in the
\mbox{$I=J=0$} channel, and so we choose a decaying exponential,
\mbox{$w(s) = \exp(-s/M^2)$}.
On the phenomenological side, this weight function will suppress the
region of the integral $s > M^2$, and so by varying $M^2$ we can
control which regions of the integrand contribute most significantly
to the sum rule.
If we also include integer powers of $s$ in the weight function then
we can build up a family of sum rules.

Almost by definition, at energies where resonances dominate the spectral
density, the implications of QCD are not straightforwardly calculable.
If QCD  \emph{were} tractable then the technology of QCD sum rules
would be redundant.
This means that, for moderate values of $s_0$, our straightforward QCD description of
$\Pi(s)$ must fail, at the very least, in the region of the positive
real axis (see Fig.~\ref{fig:pacman}).
However, it is believed~\cite{Poggio:1976} that it is \emph{only} in
the region of the positive, real axis that the straightforward 
QCD description is particularly unreliable
and that in the rest of the $s$-plane calculable QCD provides a good description
of the correlator. Consequently
we introduce a zero into the weight function at the point where the
circle joins the real axis, \emph{i.e.} $s_0$.
Thus our sum rules have the weight function  \mbox{$w(s) = s^k \left(1
- {s}/{s_0} \right) e^{-s/M^2}$} and we denote them by the symbol $T_k$.
Elsewhere in the literature they are often referred to as 'pinched
weight' sum rules. Then the left hand side of Eq.~(3) becomes
\begin{equation}
T_k(s_0,M^2)\,\equiv\,\frac{1}{\pi} \int_0^{s_0} s^k\ \left(1- \frac{s}{s_0}\right)\, \exp\left(-\frac{s}{M^2}\right) \, \Im\Pi(s) \, \dd s \quad .
\end{equation}
Introducing this zero into the weight function has a number of
advantages. 
Firstly, on the theoretical side we avoid the problem, just described,
of the failure of QCD on the part of the circle near the positive real
axis.
Secondly, on the phenomenological side, we have introduced a second
factor with a tendency to suppress the higher energy portion of the
integral and consequently reduce the dependence on the
exact value of $s_0$~\cite{LeDiberder:1992,Dominguez:1999}.

The parameter $s_0$ is not entirely free: it must be large enough that 
the QCD expressions we are able to calculate are expected to be a good 
approximation to the full correlator, yet $s_0$ should be small enough that we have
experimental information.
The art of sum-ruling is to find (sensible) values of  $s_0$ for which
\mbox{ $\left[T_k(s_0,M^2)\right]_{had} =
\left[T_k(s_0,M^2)\right]_{QCD}$} over a wide range of the unphysical 
parameter $M^2$.
When this occurs we say that the sum rules are \emph{saturated}.

Although the current~Eq.~(\ref{eq:current}) is RG-invariant, the
correlator it gives rise to is not.
As well as making the calculation of the correlator within QCD more
difficult, this would also lead to an unwanted scale dependence in our
final results.
The second derivative of this correlator, $\Pi^{\prime \prime}(s)$,
\emph{is} RG-invariant.
As $\Pi^{\prime \prime}(s)$ is also analytic within the contour shown
in Fig.~\ref{fig:pacman}, we could choose to write the sum rules
completely in terms of this quantity, but $\Im \, \Pi^{\prime
\prime}(s)$ is not directly related to any physical quantity and so we
would lose our link to experiment.
Instead we use partial integration to re-express the theoretical side of
the sum rules in terms of $\Pi^{\prime \prime}(s)$.
To do this we write the integral on the theoretical side as
\begin{equation}
i_1 = \oint v^{\prime \prime}(s) \Pi(s) \ \dd s \ .
\label{eq:intparts1}
\end{equation}
We now integrate this expression by parts twice to give,
\begin{equation}
i_1 = v^{\prime}(s) \Pi(s) - v(s) \Pi^{\prime}(s) + \oint v(s)
\Pi^{\prime \prime}(s) \dd s \ ,
\label{eq:intparts2}
\end{equation}
and arrange the constants of integration which appear in
$v(s)$ and $v^{\prime}(s)$ so that these functions disappear at
$s=s_0$.
Then our sum rule is
\begin{equation}
\frac{1}{\pi} \int_0^{s_0} w(s) \, \left[\Im\Pi(s)\right]_{had} \, \dd s =
-\frac{1}{2 \pi i} \oint_{|s|=s_0} \!\!\!\! v(s) \, \left[\Pi^{\prime
 \prime}(s)\right]_{QCD} \, \dd s \ .  
\label{eq:rgifesr}
\end{equation}

\vspace{2mm}
\baselineskip=6.0mm
\section{The Operator Product Expansion}
\label{sec:ope}

In order to calculate the vacuum polarisation within QCD we make use
of the Operator Product Expansion (OPE)~\cite{Wilson:1969}.
This replaces the matrix element of the product of currents
in~Eq.~(\ref{eq:corr}) by a linear combination of local operators ordered
in terms of increasing dimensionality,
\begin{equation}
{\Pi(s=-Q^2) = \sum_n {C_n(Q^2) \langle\mathcal{O}_n\rangle}} \ .
\label{eq:ope}
\end{equation}
The first term in this expansion is the purely perturbative
contribution, with higher dimension terms giving the contribution due
to various condensates.
We will consider terms up to dimension six, higher order terms are
expected to be negligible.
Expressions are available in the literature for the dimension zero
term to four-loop order, the dimension four term to two loops and the
dimension six term at leading order.
We quote these expressions with refernces to the original
works for the full information.
All expressions are given in the $\overline{\mathrm{MS}}$ scheme.

The dimension zero expression can be found in~\cite{Chetyrkin:1997b}
and is given by
\begin{eqnarray}
\Pi_0^{\prime \prime}(Q^2) &  = & \frac{6 m_q^2}{2 (4 \pi)^2 Q^2}
\left\{1 + a_s \left[\frac{11}{3} -2 \L \right] \right. \nonumber\\
& & \left. {}+ a_s^2 \left[ \left( \frac{5071}{144} -
\frac{35}{2}\zeta(3) \right) - \frac{139}{6} \L + \frac{17}{4} \LL
\right] \right.\nonumber \\
& & \left. {} + a_s^3 \left[ \left( -\frac{4781}{9} + \frac{b_1}{6} +
\frac{475}{4}\zeta(3)\right) + \left( -\frac{2720}{9} +
\frac{475}{4}\zeta(3) \right) \L \right. \right. \nonumber \\ 
& & \left. \left. \qquad \quad {}+ \frac{695}{8}\LL - \frac{221}{24}\LLL
\right] \right\} \ , 
\label{eq:dim0}
\end{eqnarray}
where
\begin{equation}
b_1 = \frac{4\,748\,953}{864} - \frac{91\,519}{36}\zeta(3) - 15\zeta(4)
+ \frac{715}{2}\zeta(5) \ , \nonumber
\end{equation}
and $\zeta(n)$ is the Riemann Zeta function.
As $\Pd$ is renormalisation group invariant, we are free to make any
convenient choice of the renormalisation scale.
We use the `contour improvement' prescription, whereby the logs
in~Eq.~(\ref{eq:dim0}) are `re-summed' by making the choice $\mu^2 = Q^2$.
This improves the convergence of the perturbative series by removing
the dependence on large powers of possibly large logs which would
appear in the (unknown) higher order terms.
The contour improved perturbative contribution to our correlator is
then,
\begin{eqnarray}
\Pi_0^{\prime \prime}(Q^2) = \frac{6 m_q^2(Q^2)}{2 (4 \pi)^2 Q^2} & & 
\left\{1 + \frac{11 a_s(Q^2)}{3} + a_s^2(Q^2) \left( \frac{5071}{144} -
\frac{35}{2}\zeta(3) \right) \right.\nonumber \\
& & \quad + \left. a_s^3(Q^2) \left( -\frac{4781}{9} +
\frac{b_1}{6} + \frac{475}{4}\zeta(3)\right)\right\} \ .
\label{eq:dim0Q}
\end{eqnarray}

Na\"\i vely, the only gauge-invariant operator of dimension two that can
be constructed within QCD is the so-called mass insertion term,
$m_q^2$.
For the light quarks that we are considering this term can safely be
ignored.
Some recent works have suggested that there may be mechanisms outside
of the normal OPE, which could lead to dimension two contributions to
correlation functions.
These mechanisms include renormalons~\cite{Beneke:1999} and a
tachyonic (\emph{i.e.} imaginary) gluon mass~\cite{Chetyrkin:1999}. 
It has been argued that, if they do exist, the effects of dimension two
operators would not be significant in most sum rule analyses and that
even in calculations where they would be expected to be most
noticeable, these effects are consistent with
zero~\cite{Dominguez:2000}.
Hence, in this work, we choose to ignore them.

The contour-improved contribution at dimension four can be inferred
from \cite{Jamin:1995,Surguladze:1990,Chetyrkin:1995} and is given by
\begin{eqnarray}
\Pi_4^{\prime \prime}(Q^2) & = & \frac{m_q^2(Q^2)}{2 Q^6}\left\{6 \, \mqq
\left[1 + \frac{22 a_s(Q^2)}{3} \right] - \frac{I_g}{9} \left[1 +
\frac{121 a_s(Q^2)}{18} \right] \right. \nonumber \\
& & \qquad \qquad \left. {}+ I_s \frac{4 a_s(Q^2)}{9} - \frac{3
m_s^4(Q^2)}{28 \pi^2} \right\} \ , 
\label{eq:dim4inv}
\end{eqnarray}
where
\begin{equation}
I_G = -\frac{9 \agg}{4}\left(1 + \frac{16 a_s}{9}\right) + 4 a_s \mss
\left(1 + \frac{91 a_s}{24}\right) + \frac{3 m_s^4}{4 \pi^2} \left(1 +
\frac{4 a_s}{3}\right) \ ,
\label{eq:igcps}
\end{equation}
and
\begin{equation}
I_s = \mss + \frac{3 m_s^4}{7 \pi^2} \left(\frac{1}{a_s} -
\frac{53}{24} \right) \ .
\label{eq:iscps}
\end{equation}

The leading order dimension six contribution can be taken
from~\cite{Jamin:1995} and we follow this reference in taking this
term at the fixed scale of 1~GeV.
The dimension six contribution is then
\begin{eqnarray}
\Pi_6^{\prime \prime}(Q^2) &  = & \frac{3 m_q^2(1~\GeV)}{Q^8}\left(m_q \mix
\frac{}{}\right. \nonumber \\
& & \left. {} + \pi^2 a_s \left[ \sig + \frac{2}{3} \gam \right]
\right) \ ,
\label{eq:dim6}
\end{eqnarray}
where $\gamma_{\mu}$ are the Dirac matrices, \mbox{$\sigma_{\mu \nu} = 
\tfrac{1}{2}\{\gamma_{\mu}, \gamma_{\nu}\}$} and $\lambda^a$ are the
usual Gell-Mann matrices normalised such that
\mbox{$\mathrm{Tr}[\lambda^a \lambda^b] = 2 \delta^{a b}$}. 
Currently there is no reliable method to calculate the vacuum
expectation values appearing in~Eq.~(\ref{eq:dim6}), so we follow standard
practice and relate these dimension six condensates to the (RG
variant) dimension three light quark condensate $\qq$.
For the mixed condensate, $\mix$, this is done via the
parameterisation
\begin{equation}
\mix = m_0^2 \qq \ ,
\label{eq:mix}
\end{equation}
and for the four-quark condensates, $\sig$ and $\gam$, via the
\emph{vacuum saturation hypothesis}~\cite{Shifman:1979}.
To take into account the possible violation of the vacuum saturation
hypothesis we include a multiplicative factor $V_{vs}$.
Thus our final expression for the dimension six contribution is
\begin{equation}
\Pi_6^{\prime \prime}(Q^2) = \frac{6 m_q^2}{2 Q^8} \left(m_0^2 \mqq -
\frac{176 \pi^2 a_s V_{vs}}{27} \qq^2 \right) \ .
\label{eq:dim6fin}
\end{equation}

In order to evaluate all these expressions we need to calculate the running
coupling and quark masses at complex values.
This is done by solving the four-loop Renormalisation Group Equations
that define the QCD $\beta$ and $\gamma$ functions (our conventions
for these functions are given in Appendix~\ref{app:rg}).

For the coupling, we choose to solve the implicit equation
\begin{equation}
\int_{a_s(m_{\tau}^2)}^{a_s(\mu^2)} \frac{\dd
a^{\prime}}{\beta(a^{\prime})} - \log \left( \frac{\mu^2}{m_{\tau}^2}
\right) = 0 \ ,
\label{eq:alpha}
\end{equation}
iteratively using Newton's method.  As input, we take the experimental value
$\alpha_s(m_{\tau}^2) = 0.334 \pm 0.022$, as measured by the ALEPH
collaboration~\cite{Barate:1998}.
For reference, had we used the standard expansion for $\alpha_s$, then
at four-loops and with three active flavours, this central input value
would correspond to $\Lambda_{QCD} \approx 365$~MeV.
For higher values of $\alpha_s(m_{\tau}^2)$, or correspondingly
$\Lambda_{QCD}$, the radiative corrections to~Eq.~(\ref{eq:dim0Q}) become
more important and start to swamp the higher dimension terms,
\emph{i.e} it becomes more important to include higher loop
corrections to the perturbative contribution than to include the
condensates.
It has been argued~\cite{Dominguez:2001} that the sum rule methodology
could become invalid for $\Lambda_{QCD} \gtrsim 330$~MeV.
This is especially problematic for attempts to determine condensate
values via the sum rule technique.
As this is not our aim and considering the success of previous
sum rule studies with similar and higher values of
$\alpha_s(m_{\tau}^2)$ (and $\Lambda_{QCD}$), we feel justified in applying
sum rules to this problem.

The momentum dependence of the quark mass is given by 
\begin{equation}
m_q(\mu^2) =  m_q(1 \, \GeV)\, \exp \left[\int_{a_s(1 \,
\GeV)}^{a_s(\mu^2)}\frac{\gamma(a^{\prime})}{\beta(a^{\prime})}
\, \dd a^{\prime} \right] \ .
\label{eq:mass}
\end{equation}

\section{Instantons}
\label{sec:instantons}
The OPEs for the scalar-isoscalar and scalar-isovector channels are
identical.
This was originally interpreted as an explanation for the almost exact
mass degeneracy of the $f_0(980)$ and the $a_0(980)$~\cite{Reinders:1982}.
However, we know that the OPE is not enough to  describe the
QCD vacuum completely. Other effects are important, particularly in
the scalar and pseudoscalar channels~\cite{Novikov:1981}.

The complex nature of the QCD vacuum means that it is not currently
possible to solve the equations of motion of the full
theory directly.
We must make use of approximations, which
in this work we choose to be the
Instanton Liquid
Model~\cite{Ilgenfritz:1981,Shuryak:1982,Shuryak:1982b}. 
Here the QCD vacuum is modelled as a four-dimensional `liquid' of
instantons, which is assumed to be dilute enough to make the idea of
individual instantons meaningful.
The liquid is characterised by the average instanton size $\rho_c$ and
four-volume density $n_c$ (or alternatively the average instanton
separation, $r_c = n_c^{-1/4}$).
The Instanton Liquid Model is then thought to be a good approximation,
if the ratio $\rho_c/r_c$ is small.
Within this model the instanton contribution to the
scalar-isoscalar light-quark correlator was found
in~\cite{Shuryak:1982b,Elias:1998b} to be 
\begin{equation}
\Pi_{i}(s) = \frac{3 \, Q^2 \, m_q^2}{2 \pi^2} \left[K_1(\rho_c
\sqrt{Q^2})\right]^2 \ ,
\label{eq:inscorr}
\end{equation}
where $K_1(x)$ is a modified Bessel function of the second kind (or
MacDonald function).
Taking two derivatives we obtain
\begin{equation}
\Pi_i^{\prime \prime}(s) = \frac{3 \, \rho_c^2 \, m_q^2}{4 \pi^2}
\left[K_0^{\ 2}(\rho_c \sqrt{Q^2}) + K_1^{\ 2}(\rho_c \sqrt{Q^2})\right] \ .
\label{eq:inssecdir}
\end{equation}

As stated earlier $\Pi^{\prime \prime}(s)$ is an RG-invariant quantity.
The only quantity in~Eq.~(\ref{eq:inssecdir}) that could have a
renormalisation scale dependence is the quark mass.
This renormalisation scale dependence cannot be cancelled out
anywhere else, so we conclude that the quark mass appearing
in~Eq.~(\ref{eq:inssecdir}) must be the quark mass at some fixed scale,
$\nu^2$.
\emph{A priori} we do not know what this scale is.
It is reasonable to expect that it is related to the instanton
scale given by $1/\rho_c$ or the hadronic scale of $\sim 1$ GeV, but
we have no physical reason for choosing  $1/\rho_c$ rather than
\emph{e.g.} $2/\rho_c$.
Thus we take this scale as an input parameter in the calculation.
If this scale is found to be considerably different from the hadronic
and instanton scales, then it would call into doubt our treatment of
the instanton contribution.

\baselineskip=6.2mm
\section{The Phenomenological Side}
\label{sec:phenom}

\subsection{The Coupling Schemes}
\label{subsec:ansatz}
At low energies the only hadronic process possible in the
scalar-isoscalar channel is $\pi \pi$ scattering.
As pions are made of non-strange quarks, it is reasonable to expect
that, in the elastic region, the spectral density for the correlator
is related to the absorptive part of the $\pi \pi \to \pi \pi$
scattering amplitude, $\T(s)$.
This link can be made via the scalar form-factor of the pion, $d(s)$,
which is defined by
\begin{equation}
d(s) = \langle 0|m_q \overline{q} q|\pi \pi \rangle \ .
\label{eq:formfac}
\end{equation}
Inserting a complete set of hadronic states into the
correlator~Eq.~(\ref{eq:corr}) and keeping only the lowest state,
\emph{i.e.} two pions, gives the spectral density in the
elastic region,
\begin{equation}
\frac{\Im \Pi(s)}{\pi} = \frac{3 \beta(s)}{32 \pi^2} |d(s)|^2 \ ,
\label{eq:spec}
\end{equation}
where $\beta(s) = \sqrt{1 - 4 m_{\pi}^2/s}$.

Watson's Final State Interaction Theorem~\cite{Watson:1952} implies that the
form factor and the scattering amplitude have the same phase.
Thus, in the region where elastic unitarity holds, we can write
\begin{equation}
\T(s) = \alpha(s) d(s) \ ,
\label{eq:watson}
\end{equation}
where the coupling function, $\alpha(s)$, is real for real $s > 4
m_{\pi}^2$.
Substituting~Eq.~(\ref{eq:watson}) into~Eq.~(\ref{eq:spec}) and making use of
the elastic unitarity relationship \mbox{$\Im \T = \beta \, | \T |^2 $},
we obtain 
\begin{equation}
\frac{\Im \Pi(s)}{\pi} = \frac{3}{32 \pi^2} \frac{\Im \T}{\alpha^2(s)} 
\ .
\label{eq:spec2}
\end{equation}

It is well known that $\pi \pi$ scattering contains an Adler
zero~\cite{Adler:1965}, and that this zero does not appear in
the scalar pion form factor. 
We now make the simplifying assumption that this is the only
$s$-dependence appearing in $\alpha(s)$, \emph{i.e.}
\begin{equation}
\alpha(s) = \alpha_0 (s - s_A)
\label{eq:Adler}
\end{equation}
where $s_A$ is the position of the Adler zero and $\alpha_0$ is a constant.
Substituting Eq.~(\ref{eq:Adler}) into Eq.~(\ref{eq:spec2}) gives our first ansatz
\begin{equation}
\frac{\Im\Pi(s)}{\pi} = f\,\frac{\Im \T(s)}{(s-s_A)^2} \ ,
\label{eq:ansatz1}
\end{equation}
where $f$ is an unknown constant of proportionality.

Although the $4 \pi$ threshold is at 560~MeV, it is well known that
multi-pion channels are not significant below around 1400~MeV, when quasi
two-body channels, \emph{e.g.} $\sigma \sigma$, $\rho \rho$, become
important~\cite{Bugg:1995,Abele:1996}.
The first important inelastic channel is $\pi \pi \to K
\overline{K}$.
Even after this has opened the $\pi \pi$ final state will continue to
pick out the non-strange contribution to the sum over states.
Thus we might expect Eq.~(\ref{eq:ansatz1}) to give a reasonable
approximation to the true spectral density even above 1~GeV.
We call this Coupling Scheme I.

The multi-pion final states mentioned above will increase the $n
\overline{n}$ spectral density, causing~Eq.~(\ref{eq:ansatz1}) to be an
underestimate at higher energies.
Phenomenologically, we might expect the true spectral density to be
enhanced by the ratio of the total to elastic $\pi \pi$
cross-sections, thus giving
\begin{equation}
\frac{\Im\Pi(s)}{\pi} = f\,\frac{(\Im \T(s))^2}{\beta(s) (s-s_A)^2 
|\T(s)|^2} \; ,
\label{eq:ansatz2}
\end{equation}
which we call Coupling Scheme II.
In the elastic region, the two coupling schemes are, of course,
identical.
Above the $K \overline{K}$ threshold, whereas Coupling Scheme I
ignores all inelastic channels and so will underestimate the spectral
density, Coupling Scheme II takes into account all inelastic channels,
including those with hidden strangeness, and so may be an
overestimate.

\begin{figure}[ht]
\begin{center}
\includegraphics[height=0.45\textheight]{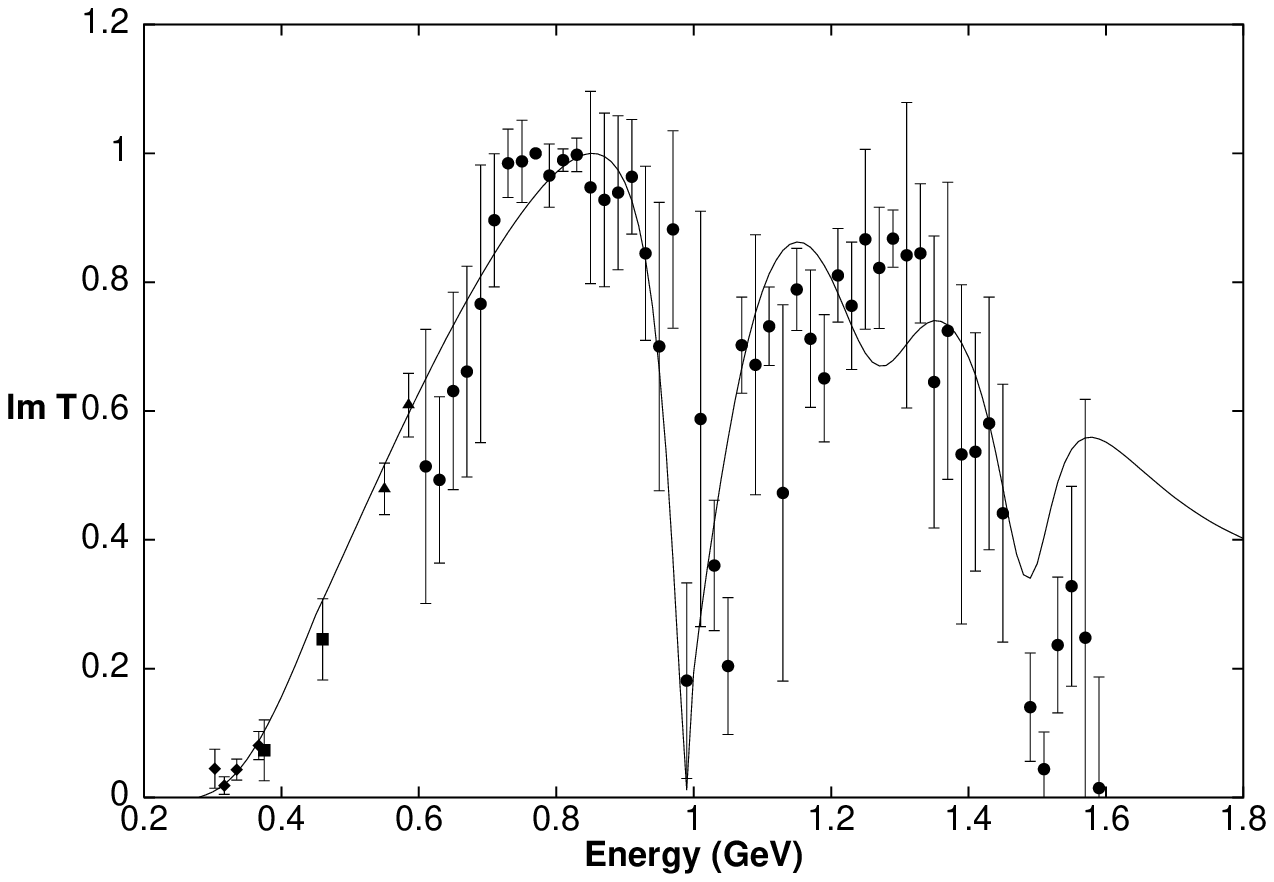}
\vspace{1mm}
\captive{A sketch showing the absorptive part of the $I=J=0$ $\pi \pi$
scattering amplitude as calculated from the data set described in the
text (line).  Superimposed are experimental points from \cite{Rosselet:1977}
(diamonds), \cite{Alekseeva:1982} (boxes), \cite{Estabrooks:1972}
(triangles) and \cite{Kaminski:1997} (circles).
\label{fig:sketch}}
\end{center}
\end{figure}

As input we take the parameterisation of the $I=J=0~\pi \pi$
scattering data carried out by Bugg~\emph{et al}~\cite{Bugg:1996}.
However, when continued down to threshold this  fails
to reproduce the scattering length predicted by Chiral Perturbation
Theory~($\chi$PT) and so in the region below 450~MeV we use the Roy
equation extrapolation of the classic Ochs-Wagner
phase-shifts~\cite{Hyams:1973}, as carried out by Pennington and used in the
discussion of $\gamma \gamma \to \pi \pi$~\cite{Pennington:1992}.

Fig.~\ref{fig:sketch} shows a sketch of the absorptive part of the
isospin zero, $S$-wave amplitude for $\pi \pi$ scattering as
calculated from our data set.
From this we can clearly see that the $f_0(400-1200)$ and the
$f_0(1370)$ appear as peaks, while the $f_0(980)$ and the $f_0(1500)$
show up as sharp dips.
Thus these latter two states, and in particular the $f_0(980)$, will 
largely decouple from the spectral density and if saturation of the
sum rules is achieved with either of our two Coupling Schemes, it will
not be due in any great part to the $f_0(980)$ or $f_0(1500)$.
By varying both $M^2$ and $s_0$ it may be possible to determine which of
the $f_0(400-1200)$ and the $f_0(1370)$ plays the dominant role in
saturation, but for the evaluation of the Wilson coefficients in the
OPE to be trustworthy, we would expect $s_0$ to be too high to exclude
the $f_0(1370)$.
Nevertheless, as the decaying exponential in the weight functions of
our sum rules suppresses the integrand for $s > M^2$, it is still
possible to determine which resonance is most important for saturation.

\newpage
\subsection{Normalisation}
\label{subsec:normal}

A value for the normalisation factor $f$, can be determined by making a
comparison with the spectral density obtained using an Omn\`es
representation~\cite{Omnes:1958} of the form factor.
This representation is given by
\begin{equation}
d(s) = d(0)\, \exp \left[ \frac{s}{\pi} \int_{4 m_{\pi}^2}^{\infty}
\frac{\phi (s^{\prime}) \, \dd s^{\prime}}{s^{\prime} (s^{\prime} -
s)} \right]\ ,
\label{eq:omnes}
\end{equation}
where $\phi(s)$ is the phase of $I=J=0 \ \pi \pi$ scattering.
The form factor at zero momentum transfer, which normalises the
Omn\`es function, can be determined from the Feynman-Hellman
Theorem~\cite{Hellman:1937} and is given by
\begin{equation}
d(0) = m_q\, \frac{\partial m_{\pi}^2}{\partial m_q} \ .
\label{eq:feynhell}
\end{equation}
To leading order in $\chi$PT, the physical pion mass is given by the
Gell-Mann-Oakes-Renner relation, \mbox{$4 \mqq = - m_{\pi}^2
f_{\pi}^2$}~\cite{Gell-Mann:1968}, and thus we see that
\begin{equation}
d(0) = m_{\pi}^2 \ .
\label{eq:dzero}
\end{equation}
Going to next-to-leading order in $\chi$PT would introduce a
correction to~Eq.~(\ref{eq:dzero}) of the order of 1\% (see for
example~\cite{Colangelo:2001}), which is safe to neglect.

To evaluate the integral in~Eq.~(\ref{eq:omnes}) exactly requires
knowledge of the scattering phase in the high energy region where it
is unknown.
However, the integrand is dominated by the region around $s$.
So if we restrict our comparison to the low energy region then the
behaviour of the phase above some point $s_1$, where $s_1 \gg s$, will
not greatly affect the value of the integral.
We take $s_1$ to correspond to the last data point and approximate the
phase above this point as a constant,~$\phi_1$.
Thus, the spectral density, as calculated from the Omn\`es
representation, is written
\begin{equation}
\frac{\Im \Pi(s)}{\pi} = \frac{3 m_{\pi}^4}{32 \pi^2}\, \beta(s) \left(
\frac{s_1}{s_1 - s}\right)^{2 \phi_1/\pi} \exp \left[\frac{2 s}{\pi} \Pint_{4
m_{\pi}^2}^{s_1} \frac{\phi(s^{\prime}) \dd s^{\prime}}{s^{\prime}
(s^{\prime} - s)}\right] \ .
\label{eq:specomnes}
\end{equation}
where the bar on the integral sign denotes the Cauchy Principal Value.
\begin{figure}[htb]
\begin{center}
\includegraphics[height=0.36\textheight]{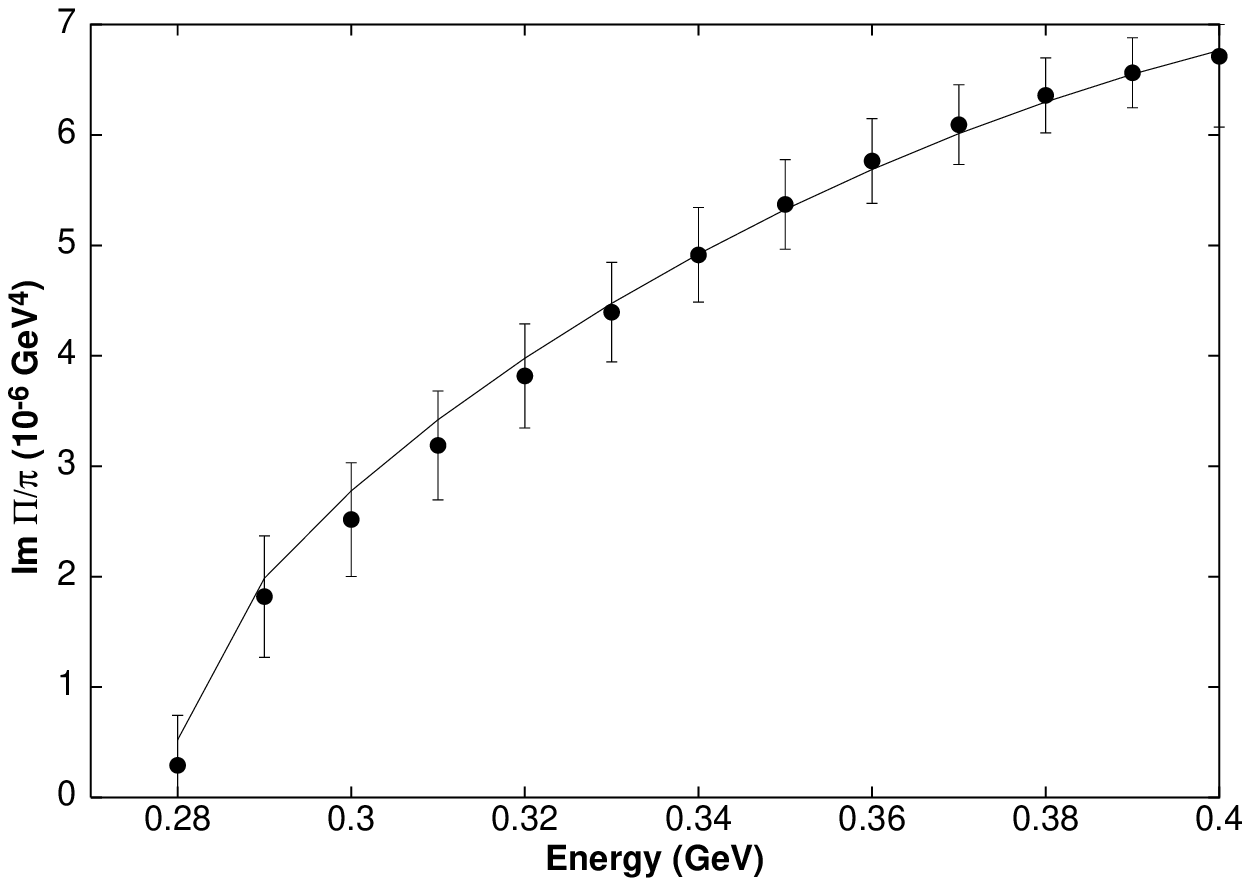}
\captive{The spectral density below 400~MeV as calculated from our
Coupling Scheme (points) and the single channel Omn\`es representation (solid line).
\label{fig:lowrho}}
\end{center}
\end{figure}

The value of the constant, $f$, can then be determined by
fitting~Eq.~(\ref{eq:ansatz1}) to~Eq.~(\ref{eq:specomnes}).
The position of the Adler zero for isospin-0, S-wave $\pi \pi$
scattering is fixed to be close to $m_{\pi}^2/2$, which is its position 
at lowest order in $\chi$PT and
so we essentially only have one parameter, $f$, to determine.
The fit, shown in Fig.~\ref{fig:lowrho}, was carried out up to a
maximum energy of 400~MeV. Over this energy range it does not matter
which Coupling Scheme is used. Each gives the value \mbox{$f = 6.9
\times 10^{-7}~\mathrm{GeV}^8$}.
However, this value is quite sensitive to the maximum energy of the
fit and should not be trusted to an accuracy better than 20\%.

\baselineskip=6.2mm

\section{Results}
\label{sec:results}

In any sum rule calculation, we hope to find regions of the parameter
space where the theoretical and phenomenological sides are equal.
We measure this \emph{saturation} of the sum rules by introducing the
double ratios
\begin{equation}
D_k(s_0,M^2) = \frac{\left[T_{k+1} / T_k \right]_{had}}{\left[T_{k+1} /
T_k \right]_{QCD}} \ .
\label{eq:drat}
\end{equation}
This ratio is independent of the overall normalising constants on
either side, \emph{i.e} $f$ and $m_q^2 (1~\GeV)$, and
obviously is equal to one when the sum rules are saturated.
As sum rule analyses are generally accurate to around 10-20\% we set an
agreement region of $(1 \pm 0.1)$.
In practice we shall use only the lowest such double ratio for each
type of sum rule, \emph{i.e} $D_0$, as the integer powers of $s$ in
the higher sum rules increase the importance of the upper end of the
integrand on the phenomenological side.

\baselineskip=5.9mm
\begin{table}[t]
\begin{center}
\begin{tabular}{|c|c|c|}
\hline
Parameter & Value & \\
\hline
$\alpha_s(m_{\tau}^2)$ & 0.334 & From Experiment~\cite{Barate:1998}\\
$\mqq $ & $ -(95 \mathrm{MeV})^4$ & From GMOR
relation~\cite{Gell-Mann:1968}\\ 
$ \agg $ & $(381~\textrm{MeV})^4$ & \\
$ \mss $ & $-(195~\mathrm{MeV})^4$ &\\
$ m_s(1~\GeV) $  & 159~MeV &\\
$ \qq $ & $-(225~\mathrm{MeV})^3$ &\\ 
$m_0^{\ 2}$ & $0.8~\GeV$ & Sum rule Estimate~\cite{Ovchinnikov:1988}\\
$V_{v s}$  & 1 &\\
$ \rho_c $ & $(600~\mathrm{MeV})^{-1}$ & \\
$\nu$ & $\sim 1 {\rm GeV}$ & To be determined\\
\hline
\end{tabular}
\captive{Standard values of all parameters used to evaluate the sum
rules.
\label{tab:parameters}}
\end{center}
\vspace{-5mm}
\end{table}

Our sum rules contain a number of input parameters which we list in
Table~\ref{tab:parameters} along with the values used.
Any deviations from these values will be discussed in
Sects.~\ref{subsec:inst}, \ref{subsec:sat} and \ref{subsec:mass}.
Where appropriate we also include a brief description of where these
values come from.

\subsection{The Instanton Parameters}
\label{subsec:inst}

The first task is to determine $\nu^2$, the scale of the quark mass in 
the instanton contribution.
To evaluate~Eq.~(\ref{eq:inssecdir}), we introduce a new
parameter,~$\lambda$, which is equal to the ratio of $m_q^2 (\nu^2)$ to
$m_q^2 (1~\GeV)$, \emph{i.e.},
\begin{equation}
\lambda = \left[  \frac{m_q(\nu^2)}{m_q(1~\GeV)} \right]^2 
= \exp \left[ 2 \int_{a_s(1 \, \GeV)}^{a_s(\nu^2)}
\frac{\gamma(a^{\prime})}{\beta(a^{\prime})} \, \dd a^{\prime}
\right] \ , 
\label{eq:lambda}
\end{equation}
where the second equality follows from~Eq.~(\ref{eq:mass}).
\begin{figure}[hbt]
\begin{center}
\includegraphics[height=0.4\textheight]{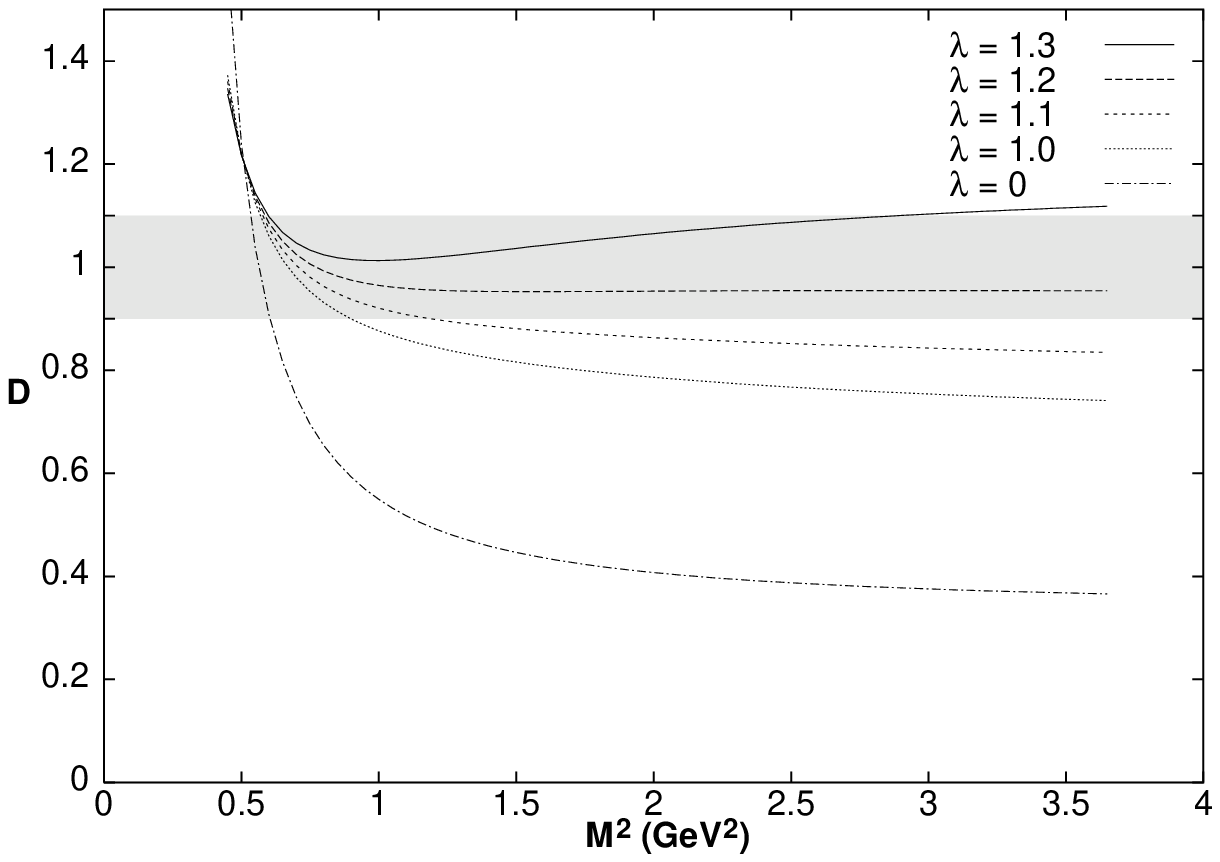}
\captive{The saturation curves obtained for various values of
$\lambda$, with $\rho_c = (480~\mathrm{MeV})^{-1}$ and all other
parameters as listed in Table~\ref{tab:parameters}.
\label{fig:fixlam}}
\end{center}
\vspace{-4mm}
\end{figure}
In Fig.~\ref{fig:fixlam} we show the stability curves obtained for
various values of this parameter using Coupling Scheme II and $s_0=3.7~\GeV$.
We have found that the saturation of the sum rules is practically
independent of $s_0$ in  the region $3~\GeV \lesssim s_0 \lesssim
4~\GeV$ and so no significance should be attached to this choice of
$s_0$.
However, it was found that saturation could not be achieved with the
standard value of~$\rho_c = (600~\mathrm{MeV})^{-1}$, and in
Fig.~\ref{fig:fixlam} we pre-empt the results of
Figs.~\ref{fig:fixrho} and~\ref{fig:fixrhofine} by using the
non-standard value $\rho_c = (480~\mathrm{MeV})^{-1}$.

It is immediately clear that the sum rules are far from satisfied when
$\lambda = 0$, \emph{i.e.} when the instanton contribution is removed.
However, the sum rules are well saturated for $\lambda \approx 1.2$
and we have found that the best value is $\lambda = 1 .25$, which we
will use from now on.
This value of $\lambda$ corresponds to a scale $\nu = 0.96$~GeV.

As stated earlier saturation can only be achieved by use of 
a non standard value of $\rho_c$. In Figs.~\ref{fig:fixrho}
and~\ref{fig:fixrhofine} we show the effect that varying this
parameter has on the saturation curves.
We see that good saturation is achieved for ~$1/\rho_c = 470 -
490~\mathrm{MeV}$, with the `best' value being $\rho_c =
(480~\mathrm{MeV})^{-1} \approx 0.42$~fm.
This value is slightly larger (25\%) than the
estimate~\cite{Shuryak:1982} based on the size of the gluon condensate
which has become the standard value employed in the Instanton Liquid
Model.
However, it is entirely consistent with current lattice determinations
which tend to find sizes in the range  0.32 -- 0.43~fm and, unless
stated otherwise, we will use this `best' value of $\rho_c$ from now on.
\begin{figure}[p]
\begin{center}
\includegraphics[height=0.42\textheight]{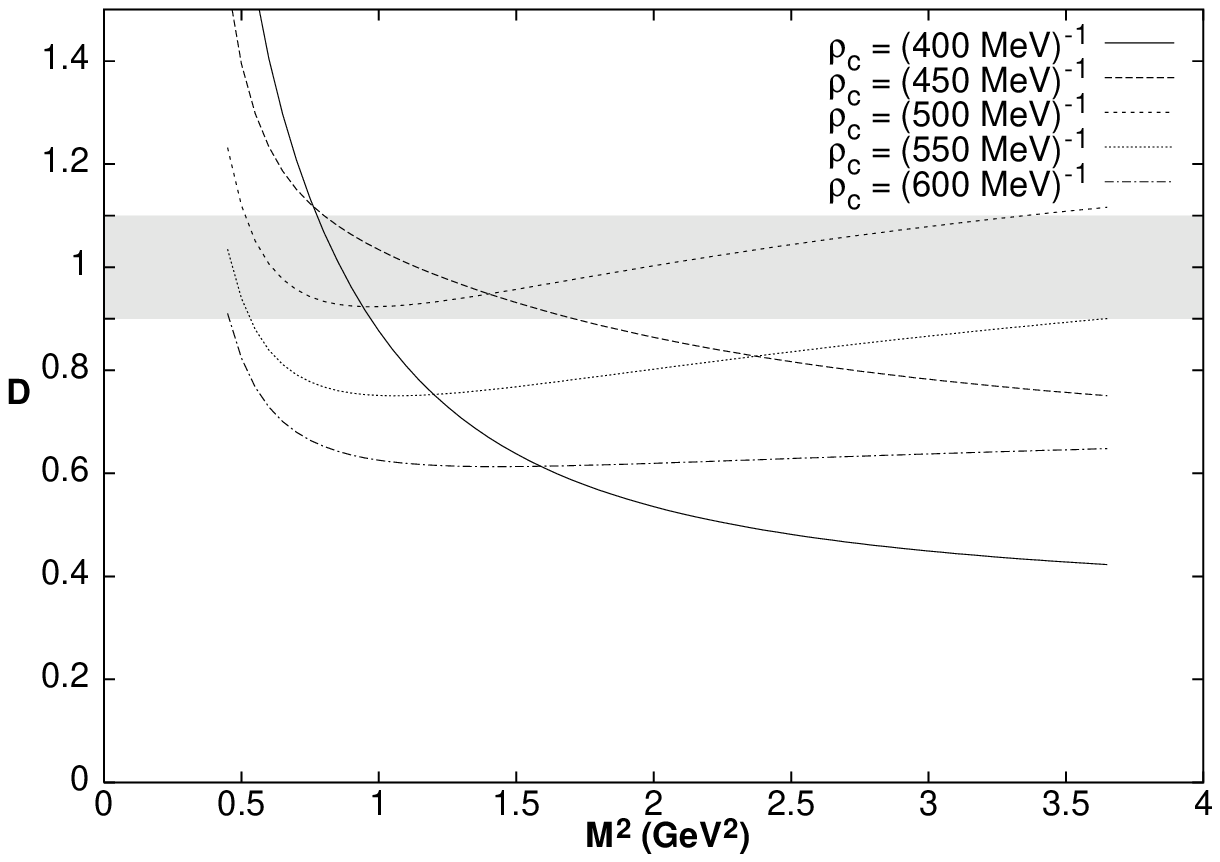}
\captive{The saturation curves obtained for various values of
$\rho_c$, with $\lambda = 1.25$ and all other parameters as listed in
Table~\ref{tab:parameters}.
\label{fig:fixrho}}
\vspace{6mm}
\includegraphics[height=0.42\textheight]{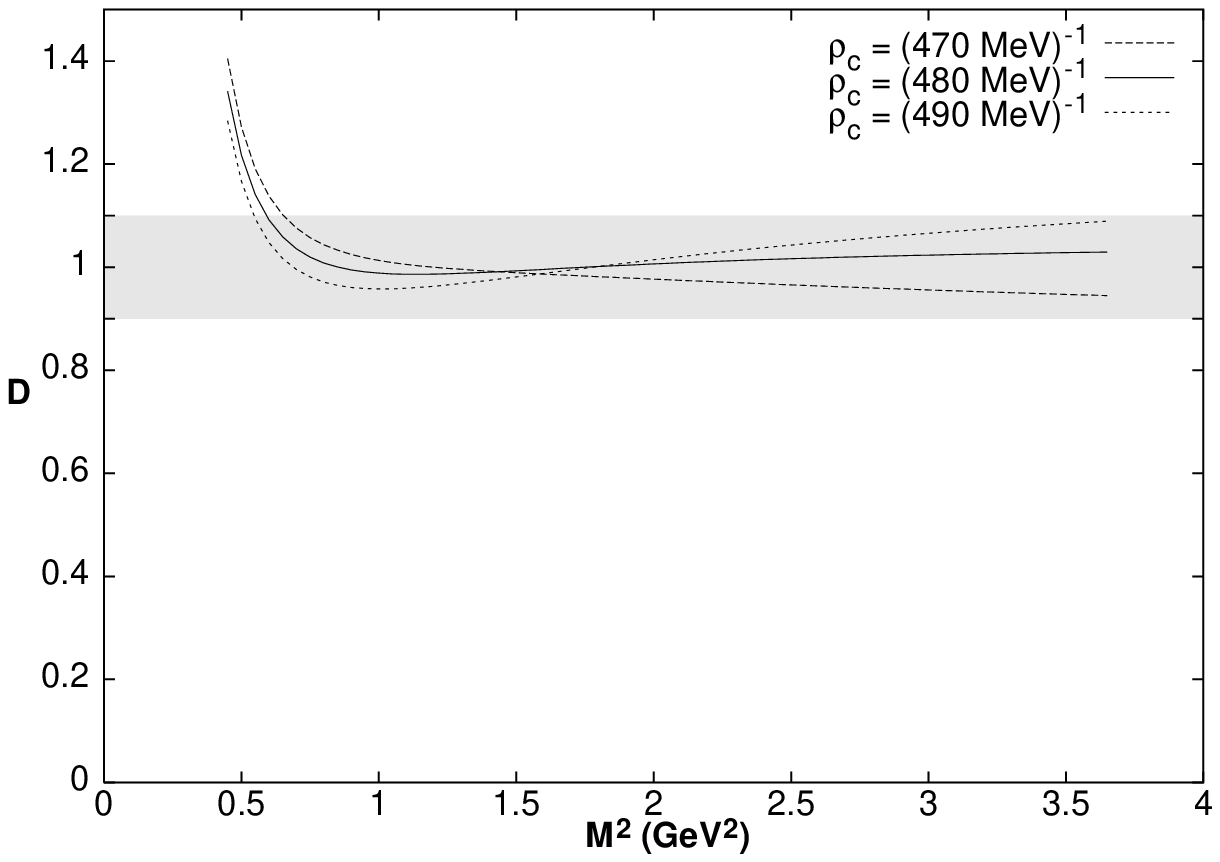}
\captive{As Fig.~\ref{fig:fixrho} but for finer steps in $\rho_c$.
\label{fig:fixrhofine}}
\end{center}
\end{figure}

\begin{figure}[p]
\begin{center}
\includegraphics[height=0.42\textheight]{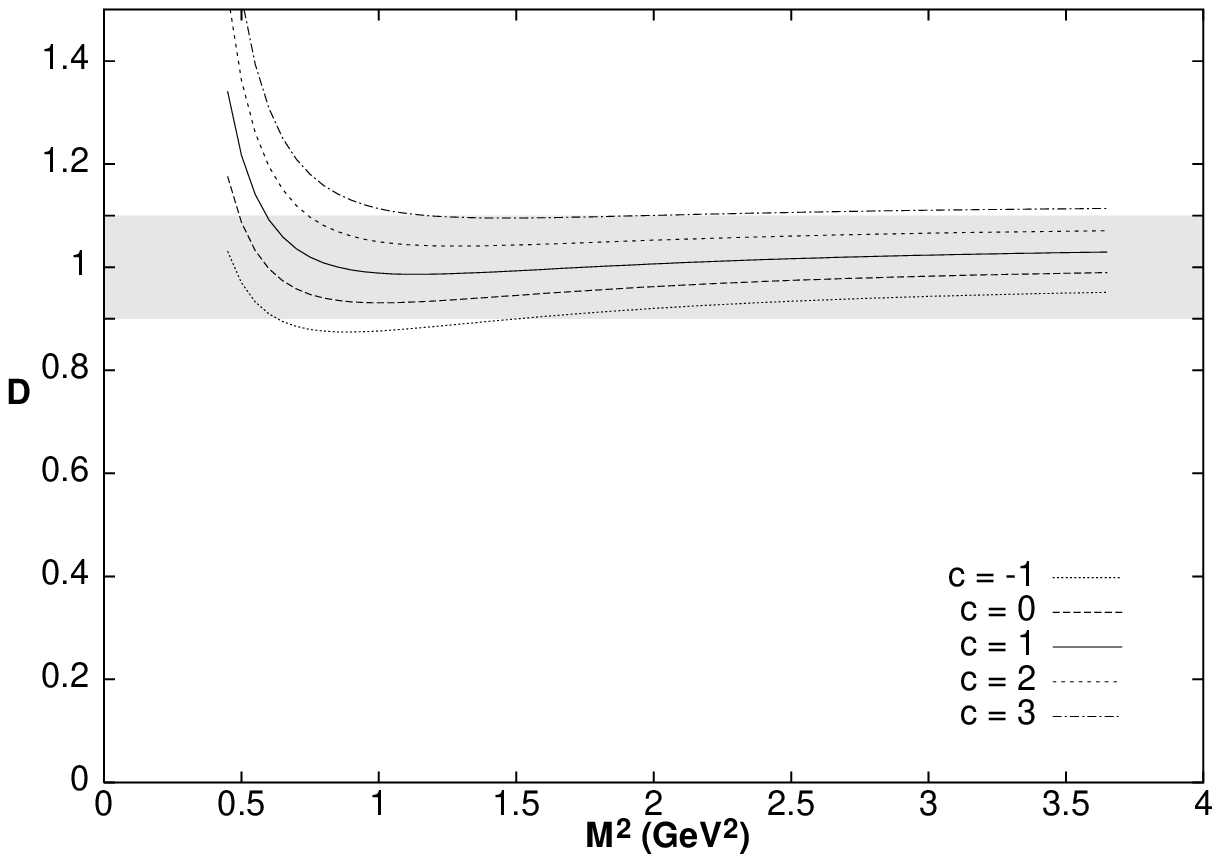}
\captive{The saturation curves, $D_0$, for various values of the
gluon condensate.
\label{fig:glusat}}
\vspace{5mm}
\includegraphics[height=0.42\textheight]{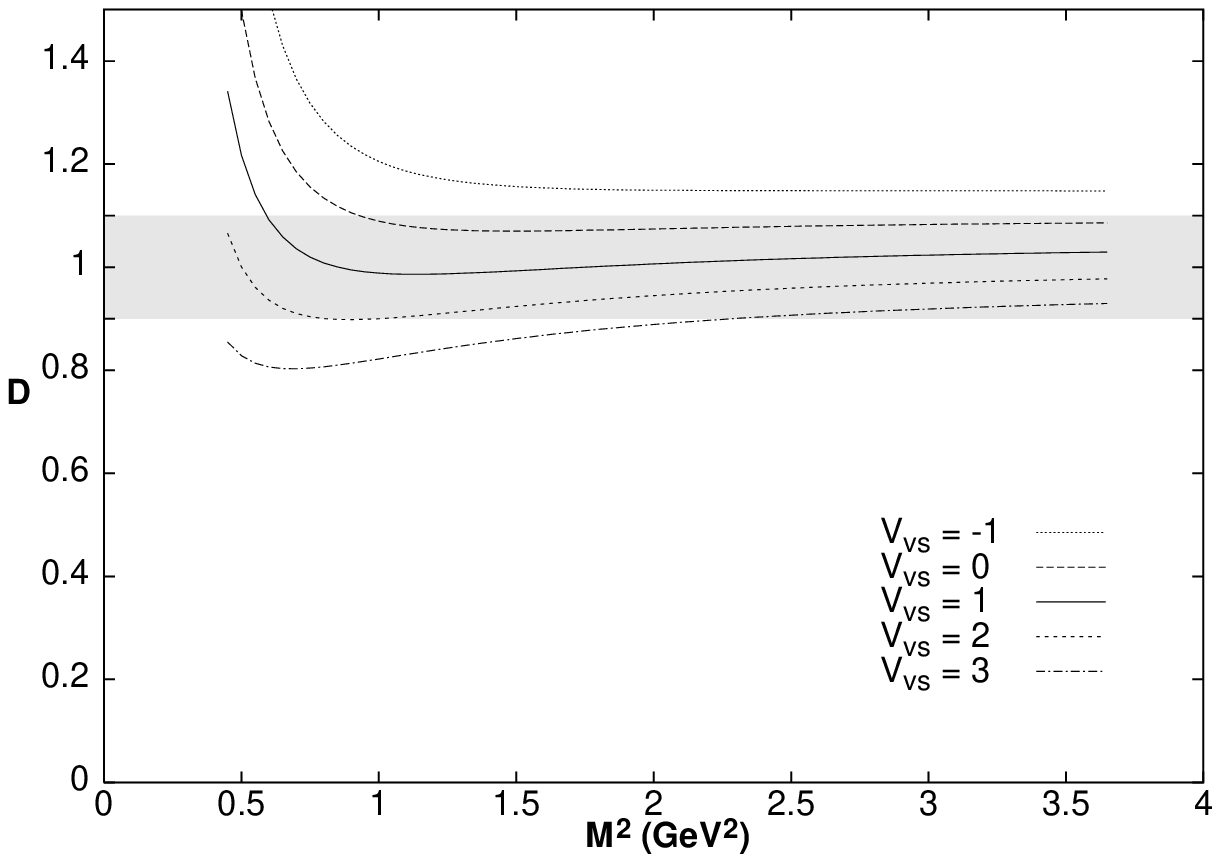}
\captive{The saturation curves, $D_0$, for various values of the
vacuum saturation violation parameter, $V_{vs}$.
\label{fig:vacsat}}
\end{center}
\end{figure}

\subsection{Saturation}
\label{subsec:sat}

As shown in Fig.~\ref{fig:fixrhofine}, once the instanton parameters 
have been determined the quality of the saturation of our sum rules is 
very good.
This means that the $f_0(980)$, being largely decoupled from our
Coupling Schemes, does not play an important role in this saturation.
On the other hand, the saturation curves are seen to be flat from quite
low values of $M^2$.
With our weight function suppressing the portion of the integrand $s >
M^2$ relative to the low energy region, the shape of the curves
implies that saturation is achieved long before the $f_0(1370)$
contributes significantly to the phenomenological integral.
This suggests that it is the $f_0(400-1200)$ that plays the most
important role in saturating our sum rules.

As the values of the condensates have not been fixed definitively, we
have performed a number of tests to determine how sensitive the
saturation of our sum rules is to variations of these input values.
To test the dependence on the size of the gluon
condensate, we write
\begin{equation}
\agg = c \agg_0
\label{eq:gcon}
\end{equation}
where $\agg_0$ is the standard value listed in
Table~\ref{tab:parameters}.
In Fig.~\ref{fig:glusat} we show the saturation curves for various
values of this factor, $c$.
We see that the saturation is not strongly dependent on the size of the 
gluon condensate, but the best results are obtained in the range  ~$0
\leq c \leq 2$.
In Fig.~\ref{fig:vacsat} we see how deviations from vacuum
saturation affect our results.
We find reasonable agreement for $0 \leq V_{vs} \leq 2$, suggesting
that vacuum saturation may be violated by 100\% with little effect.
For the remaining OPE parameters, varying their values from 0 to 200\% 
was found to have no noticeable effect.
However, the best results are obtained with the strong coupling taken
to be close to the central value quoted in Table~\ref{tab:parameters}.

\baselineskip=6.2mm

\subsection{The Average Light Quark Mass}

Another way to measure whether the sum rules are saturated is by
making a prediction for a physical quantity.
As $M^2$ is an unphysical parameter, for `sensible' values of $s_0$
this prediction should be effectively independent of $M^2$ over a wide 
range.
From~Eq.~(\ref{eq:mass}), we see that the theoretical side of our sum rules
contains an overall factor of $m_q^2 (1~\GeV)$, we can thus
make an estimate for the average light quark mass from
\begin{equation}
m_q (1~\GeV) = \sqrt{ \frac{ \left[T_k\right]_{had}}
{\left[T_k\right]_{calc}}} \ .
\label{eq:qmass}
\end{equation}
\label{subsec:mass}
\begin{figure}[hbt]
\begin{center}
\includegraphics[height=0.44\textheight]{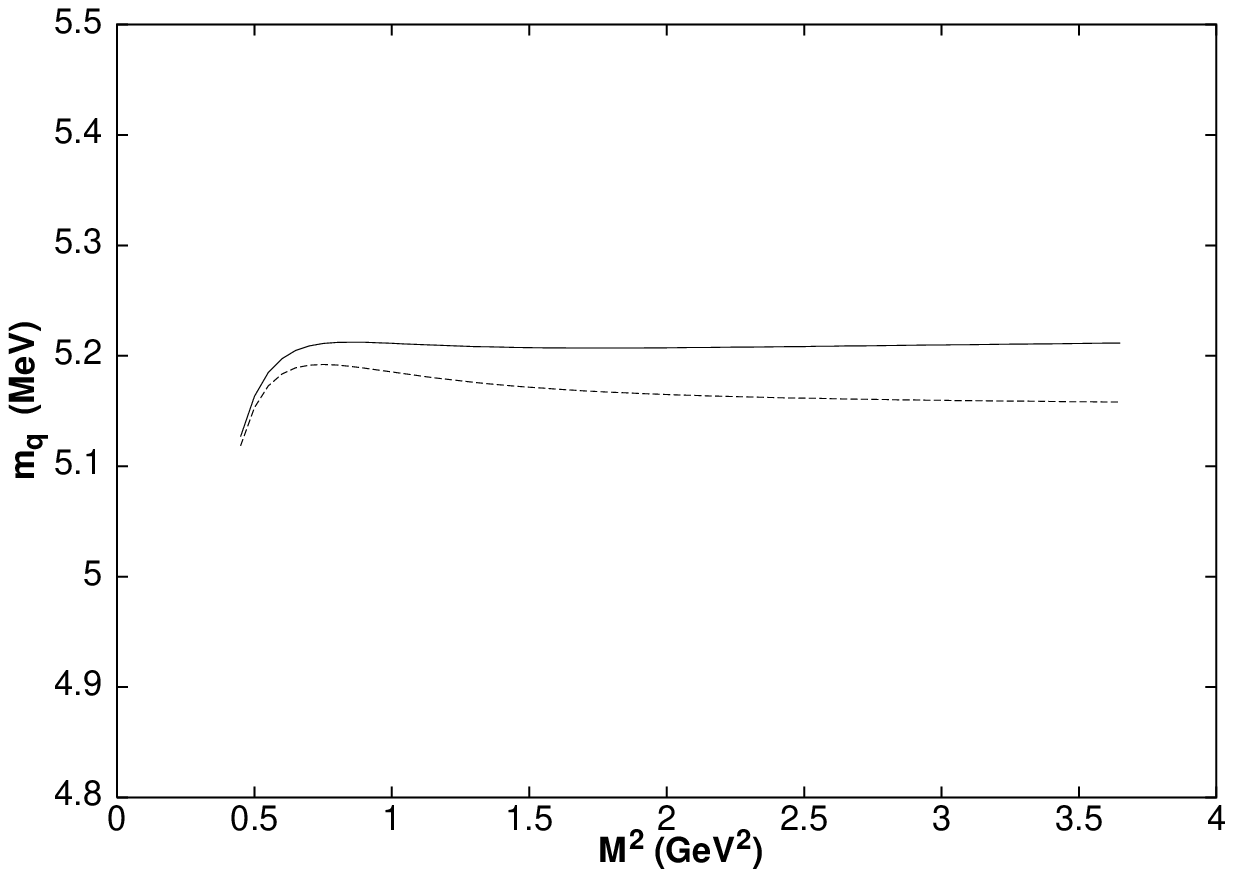}
\captive{The quark mass curves for our two Coupling Schemes.  The
dashed line shows Coupling Scheme I and the solid line corresponds to
Coupling Scheme II.
\label{fig:massscheme}}
\end{center}
\vspace{-4mm}
\end{figure}
If the graph of this quantity against $M^2$ is found to be flat then
we can say that the sum rules are saturated.
Although the actual value of the quark mass extracted will depend on
the constant $f$, its $M^2$ dependence will not, thus as a test of
saturation~Eq.~(\ref{eq:qmass}) is independent of any uncertainties in the
normalisation of the hadronic side. Once again we will use only the lowest sum rule, \emph{i.e.} $T_0$,
to estimate the quark mass.

\begin{figure}[htb]
\begin{center}
\subfigure[Average Instanton Size]{\label{fig:massrho}
\includegraphics[width=0.48\textwidth]{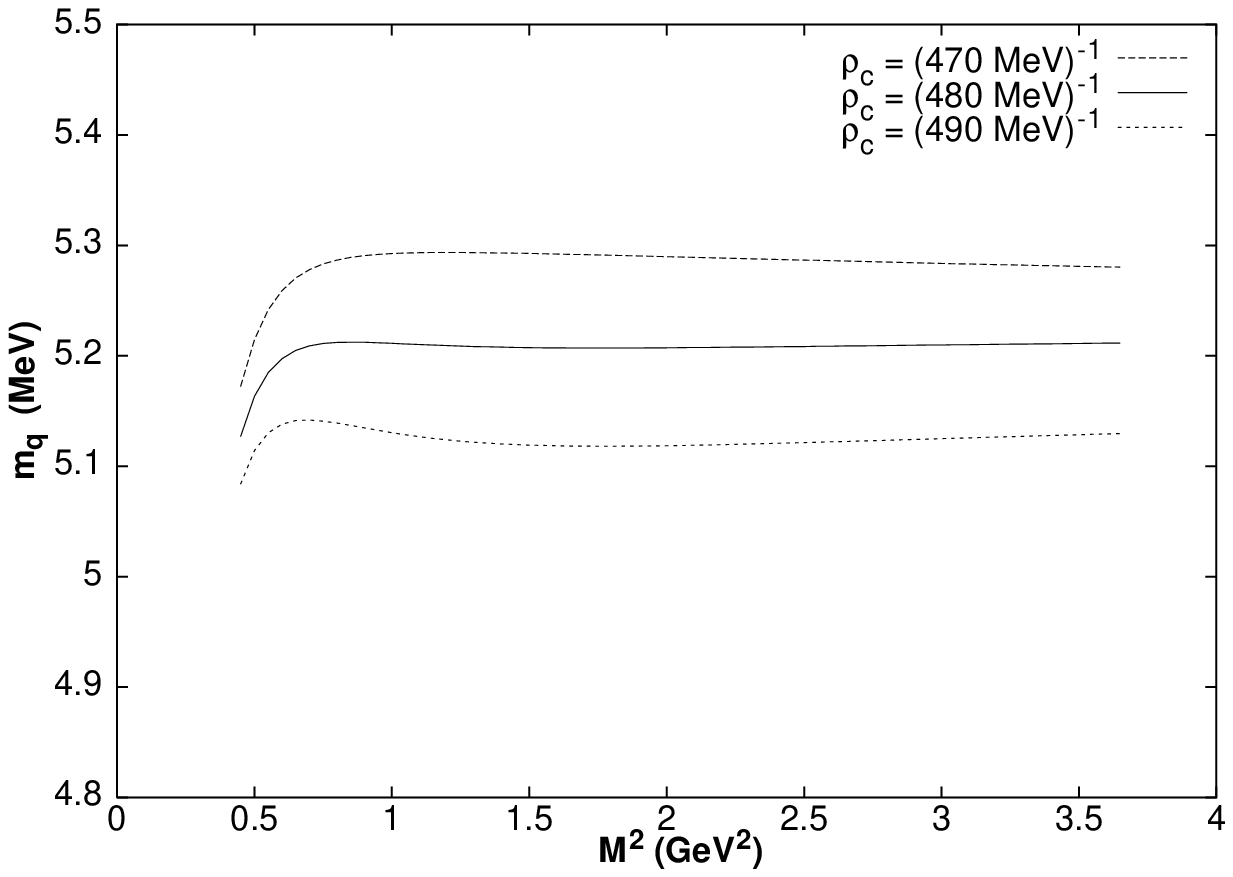}}
\subfigure[The Gluon Condensate]{\label{fig:massgluon}
\includegraphics[width=0.48\textwidth]{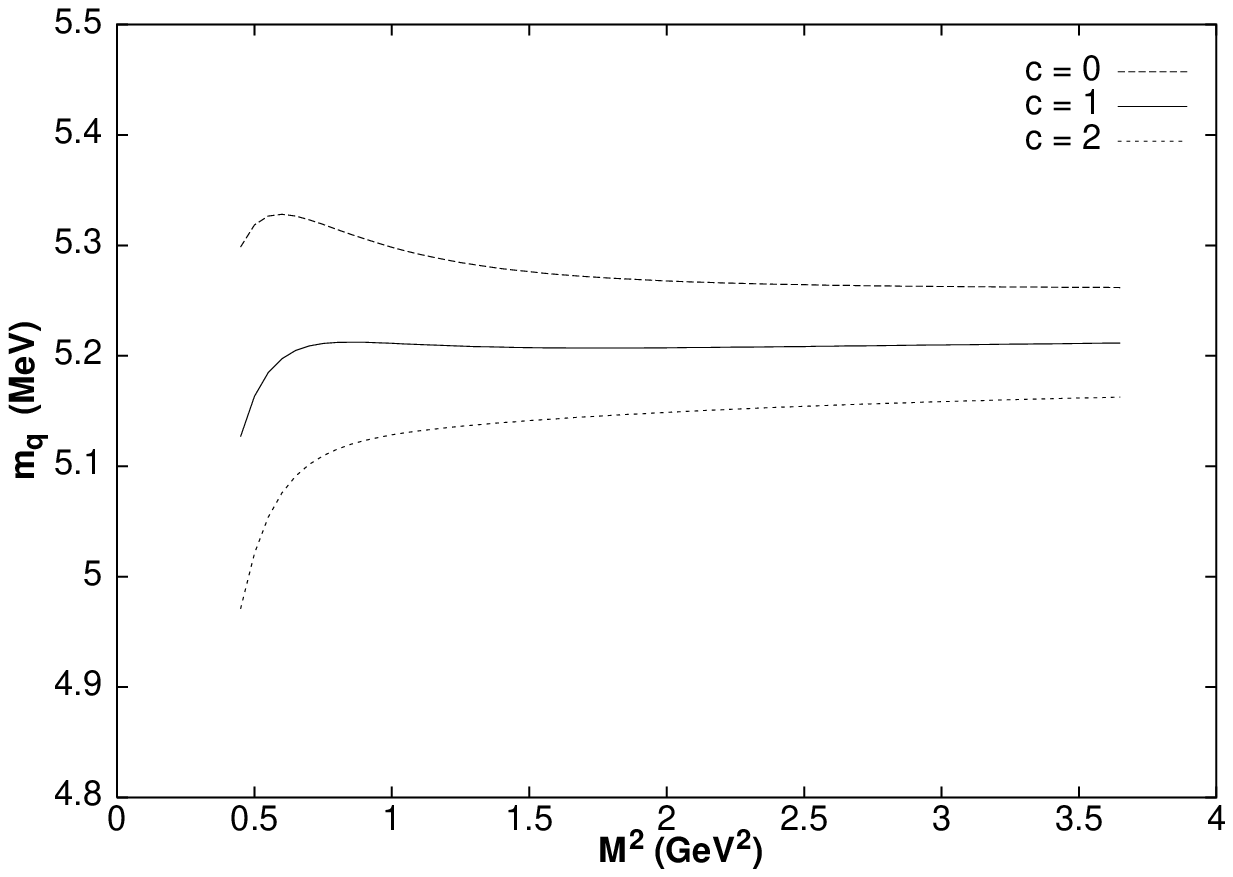}}
\subfigure[Vacuum Saturation Violation Parameter]{\label{fig:massvac}
\includegraphics[width=0.48\textwidth]{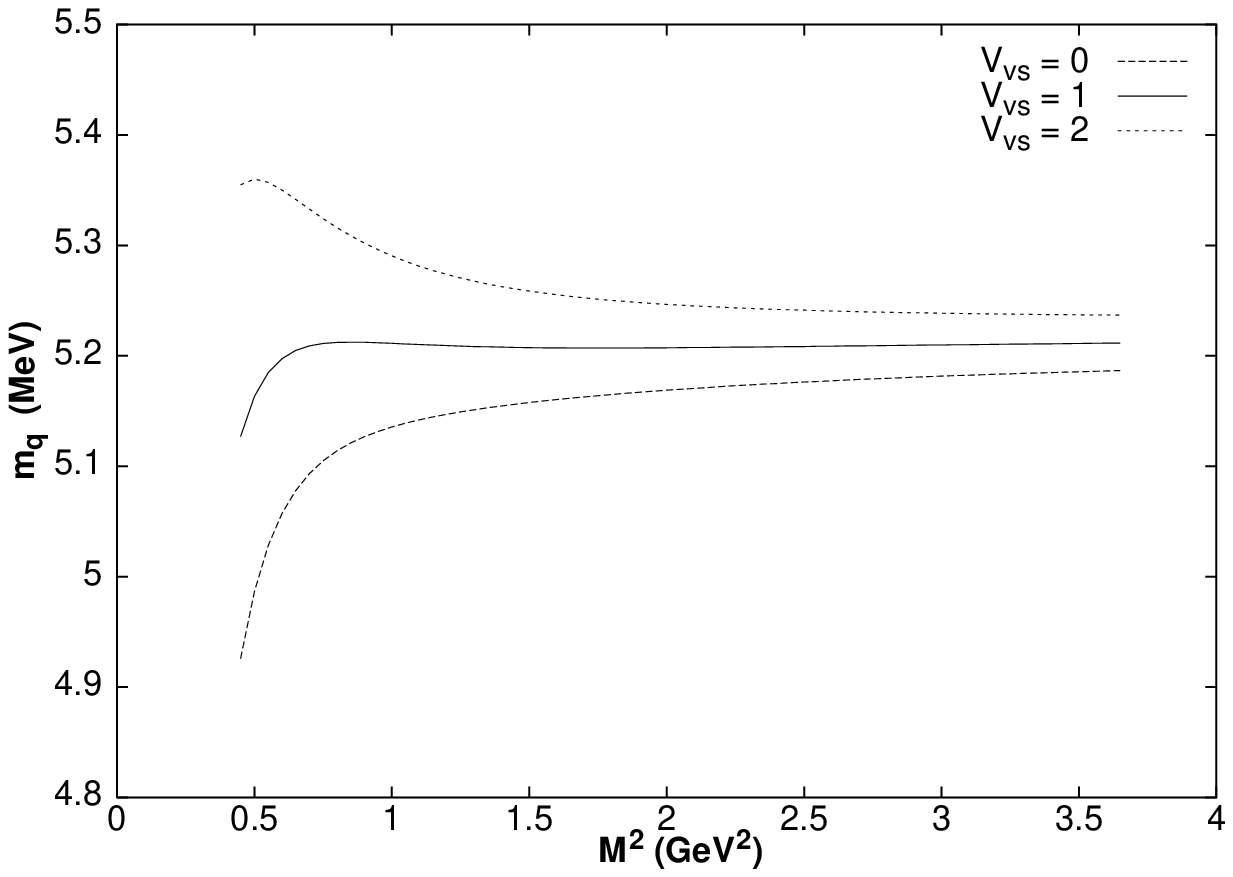}}
\caption{The effect of various parameters on the quark-mass curves.
\label{fig:mass}}
\end{center}
\end{figure}
In Fig.~\ref{fig:massscheme} we show the curve obtained
from~Eq.~(\ref{eq:qmass}) using the instanton parameters determined in
Sect.~\ref{subsec:inst} for our two Coupling Schemes.
We notice immediately that the curve is extremely flat, in particular
for Coupling Scheme II.
In both cases the variation is less than 1\% in the range $0.6~\GeV
\leq M^2 < s_0$.
This confirms that our sum rules are well satisfied even for quite low 
values of $M^2$.
The difference between the results of the two Coupling Schemes is
$\approx 1\%$.
Thus we estimate $m_q(1~\GeV) = 5.2~\mathrm{MeV}$.

\begin{table}[bt]
\begin{center}
\begin{tabular}{|c|c|c|}
\hline
& Technique & $m_q(1~\GeV)$ (MeV)\\
\hline
Chetyrkin \emph{et al.}~\cite{Chetyrkin:1999} & Pseudoscalar Sum Rule
& $5.7 \pm 1.2$\\
Prades~\cite{Prades:1998} & '' & $6.4 \pm 2.3$\\
Maltman \& Kambor~\cite{Maltman:2001} & '' & $5.6 \pm 0.8$\\
Dosch~\cite{Dosch:1998} & SR for $\qq$ + GMOR & $4.7 - 7.9$\\
APE 98~\cite{Becirevic:1998} & Quenched Lattice QCD & $6.8 \pm 0.6$\\
JLQCD 99~\cite{Aoki:1999} & '' & $6.0 \pm 0.4$\\
QCDSF 99~\cite{Gockeler:2000} & '' & $6.3 \pm 0.3$\\
APE 99~\cite{Becirevic:2000} & '' & $6.8 \pm 0.7$\\
CP-PACS 00~\cite{AliKhan:2000}& '' & $6.2 \pm 0.2$\\
SESAM 98~\cite{Eicker:1999} & Unquenched Lattice QCD & $3.9 \pm 0.2$\\
CP-PACS 00~\cite{AliKhan:2000} & '' & $4.9 \pm 0.3$\\
QCDSF - UKQCD 01~\cite{Pleiter:2001}  & '' & $5.0 \pm 0.3$\\
\hline
\end{tabular}
\vspace{2mm}
\captive{Recent determinations of $m_q(1~\GeV)$.  Note that Lattice
QCD results are generally quoted as $m_q(4~\GeV)$ and have here been
scaled up by a factor of 1.42 as required by~Eq.~(\ref{eq:mass}) with
$\alpha_s(m_{\tau}^2)=0.334$.
\label{tab:masses}}
\end{center}
\vspace{-2mm}
\end{table}

\newpage

We have performed a number of tests to determine how stable this
estimate is to the variation of input parameters within the ranges
found to give good saturation curves.
The parameters $\rho_c$, $\agg$ and $V_{vs}$ each yield an uncertainty 
in the quark mass of 0.1~MeV, see Fig.~\ref{fig:mass}.
Varying the remaining condensate values from zero to twice the value
listed in Table~\ref{tab:parameters} produces no noticeable effect on
the quark mass obtained.

The largest source of uncertainty in our calculation is the value of
the phenomenological constant of proportionality, $f$, defined in Eqs.~(27,28).
This could only be determined to an accuracy of 20\%, and so gives the dominant
uncertainty in the final result of 0.5~MeV.
Our final estimate is then $
m_q(1~\GeV) = 5.2 \pm 0.6~\mathrm{MeV}$.
In Table~\ref{tab:masses} we list some other recent determinations of
this quantity.
We see that our estimate is consistent, though lower, than earlier sum 
rule determinations.
Though not agreeing with the results of quenched lattice QCD,
our value is reassuringly consistent with the early results of unquenched lattice  calculations
which are just beginning to appear in the literature.

\baselineskip=6.2mm
\section{Discussion}
\label{sec:discuss}

The application of QCD sum rules to the isoscalar-scalar current requires
a realistic representation of the contributing resonances. None in this channel
remotely approximates a $\delta-$function, Fig.~2. Indeed, states like the
$\sigma$ (or $f_0(400-1200)$) are far too broad to be sensibly described by simple Breit-Wigner forms. Moreover they overlap with other resonances like the $f_0(980)$ and strongly coupled thresholds, like $K{\overline K}$. It is
quite natural that the $I=J=0$ correlator is expressly related to the $\pi\pi$ scattering amplitude and so the vacuum polarization can be described directly in terms of experimental information without the need for further 
(Breit-Wigner-like) approximations.
We have shown that such Coupling Schemes, defined in 
Sect.~\ref{subsec:ansatz}, are perfectly capable of saturating QCD  finite energy sum rules.
In $\pi \pi$ scattering the
$f_0(980)$ appears as a sharp dip. Consequently, in our coupling schemes this state  makes relatively little contribution to the correlator,
in total contrast to the conclusions of Shakin and Wang~\cite{Shakin:2000}.
That good saturation is possible strongly suggests then that the
$f_0(980)$ does indeed make only a small contribution to the scalar-isoscalar sum
rules, which in turn implies that it does not contain a large
$\overline{n} n$ component in its wave function.
This result is also at odds with the conclusion reached
in~\cite{Elias:1998,Shi:2000}.
However, as explained above, when the same authors~\cite{Shi:2000}
considered a more realistic approximation for the resonance shape, the
parameters found were more consistent with the
$f_0(400-1200)$ than the $f_0(980)$.
From the present analysis, beyond saying that it is not strongly
$\overline{n} n$, we can make no comment on the structure of the
$f_0(980)$.

The saturation curves we obtain are remarkably flat and close to one from low values of $M^2$, see Figs.~4-8.
As the exponential in the weight function tends to suppress the
integrand in the region $s > M^2$, saturation at these values of $M^2$
means it is the lower energy portion of the experimental
data that is most important in saturating the sum rules.
Furthermore, the $f_0(1370)$ (with mass $m$ and width $\Gamma$) will not contribute 
until \mbox{$M^2 \gtrsim (m - \Gamma/2)^2 \approx 1.4~\GeV$}.
That the sum rules are fulfilled from $M^2 \approx 1~\GeV$ implies
it is the $f_0(400-1200)$ that plays the most important role in
saturating the $\overline{n} n$ scalar-isoscalar sum rules.

Our results again show that the Operator Product Expansion is not
enough to describe the physics of the scalar mesons completely.
We find saturation only when instantons are included.
One potentially worrying point of this analysis is the need to tune the
average instanton size away from its accepted value of \mbox{$\rho_c
\approx (600~\mathrm{MeV})^{-1} \approx 0.33$~fm}.
to \mbox{$\rho_c \approx
(470 - 490~\mathrm{MeV})^{-1}$} or 0.40 -- 0.42~fm, which is about 25\%
larger.
The validity of the Instanton Liquid Model relies on the  mean
separation of instantons being significantly larger than their average 
size.
If this is not true then neighbouring instantons will overlap
significantly and it becomes meaningless to talk about individual
instantons.
The assumed average instanton separation of 1~fm is only about 2.5
times the value of $\rho_c$ required by our sum rule, and so we are
probably approaching this limit.
Reassuringly though, lattice simulations of the topological features
of the QCD vacuum seem to find average instanton sizes similar to the
one we are forced to use (for a summary of lattice determinations of
instanton parameters see~\cite{Negele:1999}).

\newpage
The above results follow from ratios of the finite energy sum rules defined in Eq.~(33) and displayed in Figs.~4-8. Moreover, the absolute value of each 
is determined by the quark mass curves of Sect.~\ref{subsec:mass}. These are also found to be
flat and this constancy sets in already below $M^2 = 1~\GeV$, Fig.~9. 
This is again evidence that our Coupling Schemes are capable of
satisfying the sum rules and that the state most important for this
is the $f_0(400-1200)$.
The value we find for the average $u$ and $d$ quark mass is
\begin{equation}
m_q(1~\GeV) = 5.2 \pm 0.6~\mathrm{MeV} \ .
\label{eq:resmass}
\end{equation}
The most conservative way to interpret this result is merely as a consistency 
check.
A number of assumptions are made in this analysis, \emph{e.g.} the
validity of the Coupling Schemes, the truncation of the OPE at
dimension-6 and the perturbative contribution at $\mathcal{O}(a_s^3)$,
the form of the instanton contribution \emph{etc}.
That our final result is of the correct size 
reassures us that none of our assumptions are too far off the mark.
Going further and taking it foregranted that the isoscalar-scalar correlator is indeed directly related to $\pi\pi$ scattering, then the sum rules provide an independent determination of the 
average light quark mass, as given in Eq.~(37).
Our result is quite compatible with sum 
rule determinations in the pseudoscalar sector and with
the early unquenched lattice estimates that
are starting to appear in the literature. We can then be certain that
 the $f_0(400-1200)$ is the dominant contribution to the $u{\overline u} + d{\overline d}$ current. 

\section*{Acknowledgements}
M.R.P. is most grateful for the warm hospitality
 as well as support at the Special Research Centre for
the Subatomic Structure of Matter at the University of Adelaide during the course of this work.
The authors would like to thank Kim Maltman 
for most helpful discussions in Adelaide and subsequent e-mail correspondence.
S.N.C. is also grateful to the theory group of the Institut de Physique Nucl\'eaire, Universit\'e Paris-Sud, Orsay for its kind hospitality.
This work was supported in part under the EU-TMR Programme, Contract
No. CT98-0169, EuroDAPHNE.

\vspace{1cm}
\appendix
\section{Renormalisation Group Conventions}
\label{app:rg}

The QCD $\beta$ and $\gamma$-functions are defined by the 
the Renormalisation Group Equations
\begin{equation}
\mu^2 \, \frac{\dd a_s}{\dd \mu^2} = - \beta(a_s) \ , \qquad
\beta(a_s) = a_s^2 \sum_{i \geq 0} \beta_i a_s^i \ ,
\label{eq:rgea}
\end{equation}
and
\begin{equation}
\mu^2 \, \frac{\dd m_q}{\dd \mu^2} = - m_q \gamma(a_s) \ , \qquad
\gamma(a_s) = a_s \sum_{i \geq 0} \gamma_i a_s^i \ ,
\label{eq:rgem}
\end{equation}
where $a_s = \alpha_s/\pi$.

Currently, the QCD $\beta$ and $\gamma$-functions are known to
four-loop order.
For three active flavours, the coefficients of the $\beta$ function
within the $\overline{\mathrm{MS}}$ scheme are~\cite{Vermaseren:1997} 
\begin{equation}
\beta_0 = \frac{9}{4} \ , \quad \beta_1 = 4 \ , \quad \beta_2 =
\frac{3863}{384} \ ,\quad \beta_3 = \frac{140\,599}{4608} +
\frac{445}{32} \zeta(3) \ .
\label{eq:beta}
\end{equation}
The coefficients of the $\gamma$-function, also within the
$\overline{\mathrm{MS}}$ scheme,
are~\cite{vanRitbergen:1997,Chetyrkin:1997c} 
\begin{equation}
\gamma_0 = 1 \ , \quad \gamma_1 = \frac{91}{24} \quad , \gamma_2 =
\frac{8885}{576} - \frac{5}{2}\zeta(3) \ , \nonumber
\end{equation}
\begin{equation}
\gamma_3 = \frac{2\,977\,517}{41\,472} - \frac{9295}{432}\zeta(3) +
\frac{135}{16}\zeta(4) - \frac{125}{12}\zeta(5) \ .
\label{eq:gamma}
\end{equation}
\vspace{6mm}


\end{document}